# Complex social contagion makes networks more vulnerable to disease outbreaks

**Ellsworth Campbell*** [1,2] **& Marcel Salathé** [1,2]

*Center for Infectious Disease Dynamics, Penn State University, University Park PA, USA*

*Department of Biology, Penn State University, USA*

***Corresponding author:***

*Ellsworth Campbell, Department of Biology, Penn State University*

*W-246C Millennium Science Complex, University Park, PA 16802, USA*

*E-mail: ells@psu.edu*

# ABSTRACT

Social network analysis is now widely used to investigate the dynamics of infectious disease spread from person to person. Vaccination dramatically disrupts the disease transmission process on a contact network, and indeed, sufficiently high vaccination rates can disrupt the process to such an extent that disease transmission on the network is effectively halted. Here, we build on mounting evidence that health behaviors - such as vaccination, and refusal thereof - can spread through social networks through a process of complex contagion that requires social reinforcement. Using network simulations that model both the health behavior and the infectious disease spread, we find that under otherwise identical conditions, the process by which the health behavior spreads has a very strong effect on disease outbreak dynamics. This variability in dynamics results from differences in the topology within susceptible communities that arise during the health behavior spreading process, which in turn depends on the topology of the overall social network. Our findings point to the importance of health behavior spread in predicting and controlling disease outbreaks.

# INTRODUCTION

Social network analysis is now widely used to investigate the dynamics of infectious disease spread from person to person, conceptualizing pathogen transmission by a diffusion process on social contact networks. A rich body of literature has explored the role of topological contact network properties such as heterogeneity in degree distributions (1, 2), cluster coefficients (3-5), and community structure (6-8) on disease dynamics. Most network-based disease dynamics models assume that everyone in the network is susceptible, and the overall contact network is taken to be the network on which the disease spreads. For many diseases, however, prior epidemics (9) and public health efforts such as vaccination (10) effectively remove individuals from the network by rendering them immune. It is therefore important to understand how these processes shape the topological properties of the network of susceptible individuals. Here, we will focus on susceptibility-promoting *behaviors*, and how the transmission of such behaviors shapes the network of susceptible individuals. We will use vaccination as a prime example of susceptibility-promoting behaviors.

Outbreaks of vaccine-preventable disease are more common when vaccination rates decline (11). High vaccination rates are therefore essential to prevent such outbreaks. In principle, partial vaccination coverage (i.e. less than 100%) can be sufficient to prevent disease outbreaks (12, 13), because a population can be protected by herd immunity if the prevalence of susceptible individuals is held below a certain threshold that depends on biological characteristics of a disease. However, outbreaks have also been observed repeatedly in countries where vaccination coverages have been increasing at already very high levels (7). For example, in 2010, many European nations reported over 10,000 measles cases while maintaining vaccination coverage rates in excess of the WHO-prescribed target of 90% (14). A growing body of research suggests that non-random distribution of unvaccinated individuals serves to counterbalance the benefits afforded by high vaccination coverage. Herd immunity is predicated on the assumption that susceptibility is spatially uniform, but geographic clustering of vaccine refusal has been widely observed (15-19). Furthermore, *Pertussis* outbreaks have been associated with the clustering of exemptions to school immunization requirements in the US (20). These studies support the supposition that communities of the intentionally unvaccinated pose a risk to local communities as well as global eradication efforts. While the causes of susceptibility clustering remain unclear, peer influence has been shown to be a significant determinant of vaccine uptake (21-23).

The spread of health behaviors such as vaccination is often modelled as a simple contagion process, similar to biological contagion, where each exposure event contributes equally to the probability of adoption of the behavior. However, there is increasing evidence (24-26) that the process of social transmission of behaviors is governed by a process of complex contagion, where social reinforcement - i.e. multiple exposures from different peers - are necessary for adoption. Here we develop a model to investigate the effects of simple and complex contagion of negative vaccination sentiment on the likelihood and size of disease outbreaks on a social contact network. Our results indicate that complex contagion increases the size of disease outbreaks, i.e. that outbreaks are largest when the spread of negative vaccination sentiment requires social reinforcement as a prerequisite to adoption. Outbreak size is further maximized when the underlying network topology is neither highly structured (e.g. lattice) nor highly unstructured (e.g. random), but rather of the "small world" type in between. We find that this is due to the interplay between the two processes, social contagion and biological contagion, and the topologies of the two networks on which these processes occur (social contagion occurs on the full network, whereas biological contagion occurs on the subnetwork of susceptible individuals only).

## METHODS

Our model is split into two time periods. In the first time period, we simulate the diffusion of negative vaccination sentiment - and subsequent vaccine refusal - on a social network. In the following, second time period, we simulate the spread of the infectious disease against which the vaccine confers complete immunity. Both processes are typically associated with considerable complexity; however, in order to keep the model tractable, we will make a few simplifying assumptions that we will outline below. We will further explore these limitations in the Discussion section.

The spread of both vaccine refusal and the infectious disease are modelled on a static social network of $N = 5000$ individuals with average degree $\bar{k} = 10$. Using the Watts-Strogatz model

(3), we model an inclusive range of network topologies depending on the rewiring probability, $p$. This model allows us to capture highly structured ring-lattices (small $p$), highly unstructured graphs (large $p$) and a variety of small world network topologies in between [Figure 1A].

Individuals participate in an opinion formation process that continues until the frequency of negative vaccination sentiment, $f_{v-}$ reaches a fixed value. The assumption of a fixed frequency of negative vaccination sentiment is in principle unrealistic, but it allows for a direct comparison of different simulation settings with identical vaccination coverage (because vaccination coverage = $1 - f_{v-}$).

The spread of negative vaccination sentiment follows a straightforward exposure - adoption process. Initially, everyone in the network has a non-negative vaccination sentiment. Once an individual's number of exposures to negative vaccination sentiment reaches a threshold, $T$, the individual adopts the negative vaccination sentiment. If $T = 1$ then the process captures simple contagion; if $T > 1$, the process captures complex contagion (26, 27). There are two ways by which an individual can be exposed to the negative vaccination sentiment: (i) the individual is exposed by a neighboring individual in the social network, or (ii) the individual is generally exposed by any other source not captured by the social contact network (e.g., through the media). In the first case of direct *social exposure*, an individual can be exposed only once by a neighboring contact. In the second case of *general exposure*, individuals can be exposed multiple times: since *general exposure* is assumed to be any other source outside of the social network, each such exposure is assumed to be from a unique source. By assuming that general exposure is ongoing at all times during the opinion formation process, we can compare three situations: general exposure only, general exposure and simple contagion, and general exposure and complex contagion [Figure 1B].

After an individual adopts the negative vaccination sentiment, neighboring contacts are exposed at rate $\Omega$ (i.e. $\Omega$ is the probability of social exposure per timestep per contact). A proportion, $r_{ge}$, of the entire social network is generally exposed to negative vaccination sentiment at each timestep. Each individual's number of unique social exposures, $e_s$, and general exposures, $e_g$, are recorded. An individual adopts the negative vaccination sentiment when their aggregate number of unique exposure events exceeds the aforementioned adoption threshold, $e_s + e_g \geq T$. Once the frequency of negative vaccination sentiment, $f_{v-}$, reaches a fixed value, all individuals with non-negative sentiments are vaccinated [Figure 1C].

After vaccination, a susceptible (i.e. non-vaccinated) individual is selected at random to seed a simple (SIR) disease epidemic. Unless noted otherwise, susceptible individuals are infected by infectious neighbors at rate $\beta = 10^{-1}$ per contact and timestep, and subsequently recover at rate $\gamma = 10^{-1}$ per timestep [Figure 1D]. Given that the average degree of the initial network is $\bar{k} = 10$, the infectious disease's basic reproductive number can be calculated as $R_0 = \frac{\beta}{\gamma} \cdot \bar{k} \cdot (1 + cv^2)$ (28) where $cv$ is the coefficient of variation of the network's degree distribution. In the case where the entire network would be susceptible to the disease, the resulting $R_0$ of ~10 would require a vaccination coverage of $1 - \frac{1}{R_0} \approx 90\%$ to provide for herd immunity, a value that was chosen to lie approximately in the middle between moderately transmissible diseases such as influenza (with an estimated $R_0$ between 1 and 3) and highly transmissible diseases such as pertussis and measles (where the latter has $R_0$ estimates in excess of 10). The infection process continues until all infected individuals have recovered. For each round of opinion formation considered, multiple independent disease epidemics are simulated. For each epidemic simulation, we record the number and size of susceptible communities, or connected components of susceptible individuals, as well as the final epidemic size. We define an outbreak as a final epidemic size larger than 25 individuals, which corresponds to 0.5% of the total population.

## RESULTS

The topological distribution of individual susceptibility on a population's contact network strongly affects the probability of a disease outbreak in that population. To begin, we will look at three different sentiment spreading scenarios (random, simple and complex) on three different types of network topologies, defined by the rewiring probability $(p)$: highly structured $(p = 0.01)$, intermediately structured, i.e. "small world" $(p = 0.1)$, and highly unstructured $(p = 0.5)$. We assume a population's baseline risk of experiencing an outbreak to be the frequency of disease outbreaks when the formation of public opinion regarding negative vaccination sentiment is a general exposure process that occurs in the absence of social contagion. The outcome of this process is equivalent to a random distribution of vaccination status on the network, and as a consequence, outbreaks are rare and approach zero as the proportion of vaccinated individuals approaches the herd immunity threshold of 90% [Figure 2A]. We then compare the results of this baseline scenario to two social spreading scenarios, simple and complex contagion. In both scenarios, sentiments predominantly spread through social exposure [Figure S1], but it is worth repeating that general exposure is ongoing at low rates. When negative vaccination sentiment spreads by simple contagion, outbreaks dramatically increase in frequency compared to the random baseline scenario, and even occur when vaccination coverage has approached the herd immunity threshold. However, at vaccination coverage of 95%, no outbreaks occur in the simple contagion scenario. The situation is different when negative vaccination sentiment spreads by complex contagion: outbreaks are generally more frequent than even under the simple contagion scenario, and importantly, they still occur at 95% vaccination coverage, an outcome not observed with simple contagion. Overall, the results are strongly dependent on the way by which negative vaccination sentiments spread.

It is important to note that the original network topology affects the disease outcome in two ways. First, the network topology will affect how negative vaccination sentiments spread. Second, once the negative vaccination sentiments have spread, the structure of the remaining smaller subnetworks of susceptible individuals - the networks on which the disease can spread - will affect disease dynamics. Thus, the network topology of the original network will also affect the topology of the emerging susceptible networks, simply because the susceptible networks are subnetworks of the original network. We will first focus on the number and size of susceptible subnetworks that are generated by negative vaccination sentiment spread, and we will refer to these subnetworks as communities (defined as a group of nodes where each node in the group is connected to each other node in the group by a path, but to no other node outside of that group - this is also known as a weakly connected component in graph theory). The size and frequency of outbreaks will be strongly affected by the number and size of susceptible communities. Second, we will focus on the topology of the susceptible communities, and its effect on disease dynamics.

Complex contagion of negative vaccination sentiment produces slightly fewer susceptible communities [Figure 2B] than simple contagion. Under simple contagion, each general exposure of a non-exposed individual leads to the adoption of the negative vaccination sentiment, at which point it can spread from this initial seed and give rise to an expanding susceptible community. Under complex contagion, both the adoption and the spread of negative vaccination sentiment proceeds more slowly because of the $T > 1$ requirement. Because we run simulations until a fixed fraction $f_{v-}$ of the population has adopted the negative vaccination sentiment, the slower spread in the complex contagion scenario means that there is more time for new communities to emerge. However, the initial generation of a novel community is such a rare event in the complex contagion scenario that it more than compensates for the effect of a longer time frame, resulting in fewer susceptible communities overall than in the simple contagion scenario. This finding is consistent over a wide range of parameters [Figure S2].

Our assumption that a fixed fraction $f_{v-}$ of the population adopts the negative vaccination sentiment means that the number of communities relates directly to the average size of these communities. In particular, since complex contagion generally produces fewer communities than simple contagion, these communities are on average larger. However, the average size alone can be a poor guide to predict final sizes of disease outbreaks because the community size distribution is often skewed, typically with one large community and a few very small ones. To capture the distribution in a single number that relates to expected outbreak size, we used a quasi-deterministic version of the model with deterministic disease transmission ($\beta = 1$) and without recovery ($\gamma = 1$). In this quasi-deterministic model, the susceptible index case is chosen randomly, after which the disease outbreak will completely saturate the community of susceptible individuals in which it was started. Such outbreaks represent an upper bound – and thus worst-case scenario – of the final size in a given community. Since there are often multiple susceptible communities, we can calculate an average upper bound, $\bar{F}$, of final outbreak size.

$$\bar{F} = \frac{\sum C_i^2}{\sum C_i}$$

Hence, $\bar{F}$ is a weighted mean size of all susceptible communities $C_i$ in the contact network. The susceptible community's size serves to weight the mean, because a randomly infected index case is more likely to be a member of larger communities. This estimate assumes that the disease epidemic will deterministically saturate the index case's community, regardless of the size and topology of the community. In the quasi-deterministic simulations, we find that complex contagion produces sets of communities that have higher upper bounds in outbreak sizes than simple contagion [Figure 2C]. This effect is particularly pronounced in more structured networks (i.e. small $p$); in more randomized networks, the average distance between any two nodes is low (3), and the set of nodes that have adopted the negative vaccination sentiment are more likely to be connected in a single component simply due to the underlying original network structure, rather than due to the effects of social contagion.

If we relax these quasi-deterministic constraints and simulate stochastic infectious disease epidemics with $\beta = 0.1$ and $\gamma = 0.1$, we find that increases in rewiring probability, $p$, are no longer predictive of increases in final epidemic size [Figure 2D]. While all vaccine-averse individuals are equally susceptible to infection, the communities to which they belong are not equally susceptible to saturation by infectious disease. This variability results from differences in the topology *within* susceptible communities, which arises during the opinion formation process. Under complex contagion, individuals can adopt negative vaccination sentiment by social contagion or general exposure as well as a combination of the two processes. As the topology of the initial contact network becomes less structured and more random (i.e. large $p$), the availability of social reinforcement decreases (25), resulting in an increased proportion of adoption events that are caused by a mix of general and social exposures [Figure S1]. Alternatively, under simple contagion, the proportion of general or social adopters depends only on the rates of general and social exposures rather than the underlying network topology. Because infectious individuals may recover before infecting a neighbor, the increased path redundancy caused by complex contagion ensures that an infected individual has ample opportunity to transmit to a susceptible neighbor before recovering, reducing the chance of stochastic fade-outs. This can be best captured by calculating the mean basic reproductive number ($R_0$) in a susceptible community - its value will depend solely on the mean degree and variance of the degree distribution, as the transmission and recovery rates are identical. As can be seen in Figure 2E, $R_0$ is greater in communities generated by complex contagion than in communities generated by simple contagion. (Note that $R_0$ is less than 10 under both simple and complex contagion because vaccinated individuals are removed by virtue of their immunity to infection, and the average degree ($\bar{k}$) of susceptible communities is less than 10.)

Taken together, the results suggest that infectious disease outbreaks are substantially larger on contact networks shaped by complex social contagion than on networks shaped by simple contagion. They also suggest the outbreaks are largest in networks whose topology is neither highly structured nor highly random, but rather best described as "small world" topologies, characteristic of many social (contact) networks (29, 30). The susceptible communities generated by social contagion in highly structured networks are smaller on average, but the resulting $R_0$ is higher in these communities - vice versa, susceptible communities generated by social contagion in highly random networks are larger on average, but the resulting $R_0$ is lower. In the parameter space of small world networks, both the average community size as well as $R_0$ are moderate, but in combination they generate the largest outbreaks in the parameter space tested.

## DISCUSSION

The primary finding of our research is that infectious disease outbreaks are larger and occur more frequently when susceptibility-inducing behaviors, such as negative vaccination sentiment, spread across contact networks by complex contagion rather than by simple contagion. Contact redundancy, or the density of social reinforcement, within a susceptible community is a strong determinant of both the size and frequency of disease outbreaks. The density of potential social reinforcement is determined by how structured or random the contact network's topology is prior to the period of opinion formation. Complex contagion of negative vaccination sentiment fosters redundancy within communities of unvaccinated individuals, resulting in susceptible communities that are more readily saturated by infectious diseases. Our results indicate that standard estimates (31) of vaccination rates to attain herd immunity can be highly insufficient to protect a community if clustering of susceptible individuals is caused by the social spread of negative vaccination sentiments, and particularly so if the contagion process is complex, requiring social reinforcement.

Given that peer influence is a significant determinant of vaccine uptake (21-23) in regions where vaccine availability is not a limiting factor, conditions found in high-income nations may serve as early indicators of future hurdles to global eradication efforts. A recent survey study of hospital workers illustrates that the expressed reasons for vaccine refusal are most strongly associated with myths and urban legends about immunization, leading to concerns about adverse effects and insufficient efficacy (32). This problem of perception is traditionally approached from a game-theoretic perspective wherein individuals are assumed to perform a complex risk-analysis with respect to financial cost, treatment efficacy, the risk of infection, etc. However, rather than disentangle such a complicated decision, individuals may defer to social reinforcement as a rough proxy for an informed cost-benefit analysis. Modeling a complex social contagion that affects disease-susceptibility allows us to underscore the role of social deferment in the adoption of health-behaviors that are both risky and beneficial.

To the best of our knowledge, this is the first study to look at the effect of complex contagion of vaccination behavior on infectious disease dynamics, and as such it is limited in several ways. We have focused on the effect of a minimal change in the adoption threshold $T$ that differentiates complex from simple contagions. There is little doubt that adoption threshold is variable between both individuals and social contagions themselves. Sociologists have long recognized the influential role of "early adopters": individuals characterized by a low adoption threshold to the contagion under investigation. A recent theoretical model (33) explores cascade dynamics when early adopters are also more active and enthusiastic spreaders. Barash et al. (34) have also considered cascade dynamics when the adoption threshold is not only variable, but also determined relative to the proportion of neighboring adopters. Both are particularly valuable lines of inquiry as we consider contagions in competition with other, mutually exclusive contagions.

Further, we do not explicitly simulate the spread of positive vaccination sentiment; rather positive sentiment is treated as a default position for individuals who do not adopt negative vaccination sentiment by the end of the opinion formation period. Oppositional social contagions (e.g., positive and negative vaccination sentiment) are often in competition for individual attention (35), but this paradigm may not apply to vaccination in developed nations because immunization is often a default prerequisite for access to public institutions. As a result, those who may hold a neutral sentiment are incentivized to vaccinate. Institutional immunization in developed nations ensures a wide and largely uniform distribution of vaccinated individuals, though it does not preclude further contagious spread of positive vaccination sentiment. Indeed, the prevalence of positive vaccination sentiment may have synergistic or antagonistic effects on the spread of negative vaccination sentiment.

Finally, we model public opinion formation and the spread of infectious disease as two serial processes that occur on the same, static contact network. While these assumptions may be justified in some circumstances, they need to be relaxed in future studies. For example, modern communication technologies and services (mobile phones, social media, etc.) can result in communication networks that can be rather divergent from the contact networks upon which infectious diseases can spread. With respect to the temporal dynamics of the two spreading processes, we recognize that public opinion about the decision to vaccinate is a continuous, dynamic process that can be affected by the global and local prevalence of infectious disease (36). Furthermore, dynamic social interactions that are not captured by static contact networks are increasingly important in the realm of highly communicable diseases such as measles (5, 10, 37). We hope that our assumptions are understood as necessary simplifications in this initial exploration that allow for the direct comparison of epidemic outcomes in susceptible communities whose only difference arises from either the simple or complex contagion of negative vaccination sentiment.

## ACKNOWLEDGEMENTS
The authors would like to thank Timo Smieszek. EC and MS acknowledge computational cluster computing resources funded by the *National Science Foundation* through grant OCI–0821527. MS acknowledges funding from a *Branco Weiss Fellowship*. EC acknowledges funding from the *University Graduate Fellowship* by the *Eberly College of Sciences*. The funders had no role in study design, data collection and analysis, decision to publish, or preparation of the manuscript.

**Figure 1**: Schematic representation of the complex contagion of negative vaccination sentiment followed by an SIR disease epidemic. White nodes denote non-adopters of a negative vaccination sentiment. Black nodes denote individuals who adopt a negative vaccination sentiment. Red nodes denote individuals who have been infected. **(A)** Initial social contact network. **(B)** After negative vaccination sentiment spreads by complex contagion during the period of opinion formation **(C)** After vaccination, and subsequent removal of immunized individuals from the susceptible contact network. **(D)** After a vaccine-preventable infectious disease spreads through the remaining susceptible network.

**Figure 2:** Estimated and simulated epidemiological measures if an infectious disease spreads through susceptible communities that are generated by the social transmission of negative vaccination sentiment. Parameter ranges for all simulations are shown. All points are averages based on 100 unique susceptible networks generated by stochastic simulations of social contagion. **(A)** Frequency at which infectious disease outbreaks occur in a population. An outbreak is defined as a minimum final epidemic size of 25 (i.e. 0.5% of the total population size N=5000). For each unique network we ran 10,000 infectious disease simulations. **(B)** Number of distinct susceptible communities that are generated by the social transmission of negative vaccination sentiment. $r_{ge} = 10^{-5}$, $\Omega = 10^{-4} \leftrightarrow 10^{-2}$, $f_{v-} = 0.10$. **(C)** Quasi-deterministic final epidemic size. Shaded region denotes 95% Confidence Intervals. $\beta = 1$, $\gamma = 0$, $r_{ge} = 10^{-5}$, $\Omega = 10^{-4} \leftrightarrow 10^{-2}$, $f_{v-} = 0.10$. **(D)** Simulated final epidemic size. Shaded region denotes 95% confidence intervals. $\beta = 10^{-1}$, $\gamma = 10^{-1}$, $r_{ge} = 10^{-5}$, $\Omega = 10^{-4} \leftrightarrow 10^{-2}$, $f_{v-} = 0.10$. For each unique network we ran 10,000 infectious disease simulations **(E)** Mean basic reproductive number ($R_0$) of an infectious biological agent in susceptible communities that were generated by the social transmission of negative vaccination sentiment. $\beta = 0.1$, $\gamma = 0.1$, $r_{ge} = 10^{-5}$, $\Omega = 10^{-4} \leftrightarrow 10^{-2}$, $f_{v-} = 0.10$

**Supplementary Figure S1**: The susceptible population, categorized by the exposure events that caused adoption of negative vaccination sentiment. 100 simulations, $r_{ge} = 10^{-5}$, $\Omega = 10^{-4} \leftrightarrow 10^{-2}$, $f_{v-} = 0.10$.

**Supplementary Figure S2**: The number of susceptible communities formed by either complex or simple contagion of negative vaccination sentiment. 100 simulations, $r_{ge} = 10^{-5} \leftrightarrow 10^{-2}$, $\Omega = 10^{-4} \leftrightarrow 10^{-2}$, $f_{v-} = 0.10$

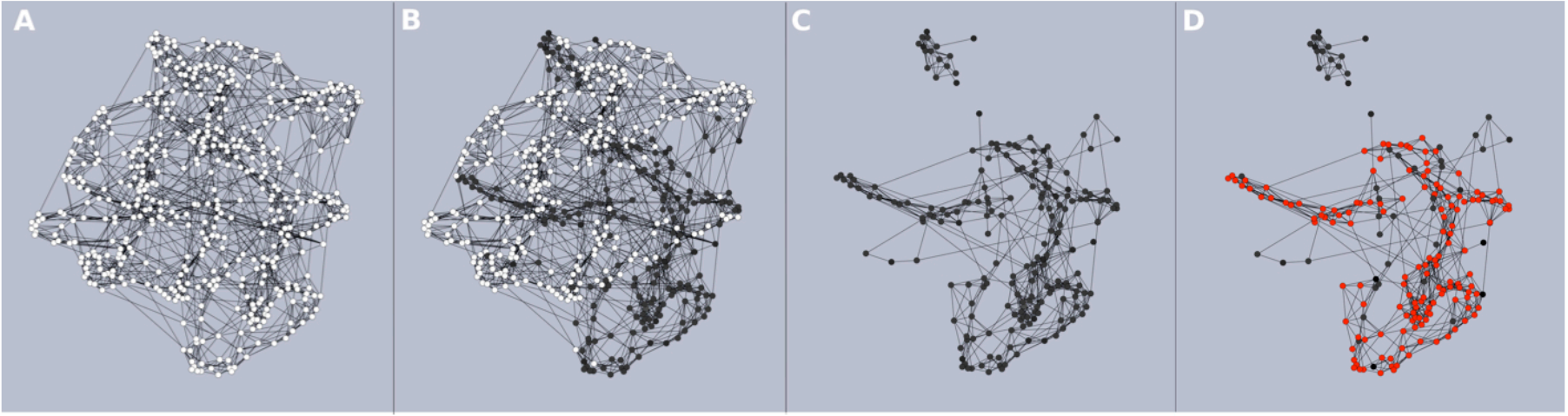

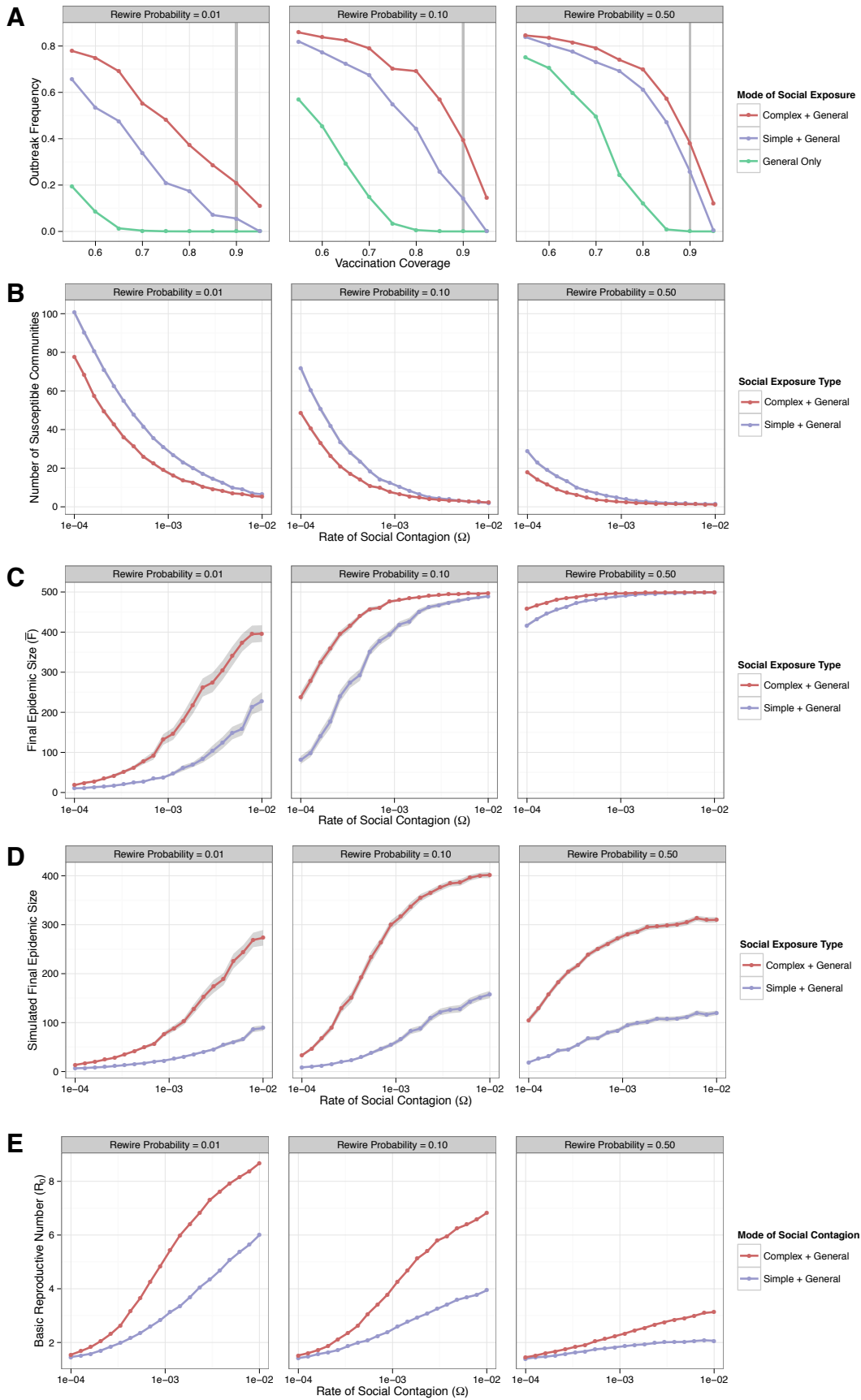

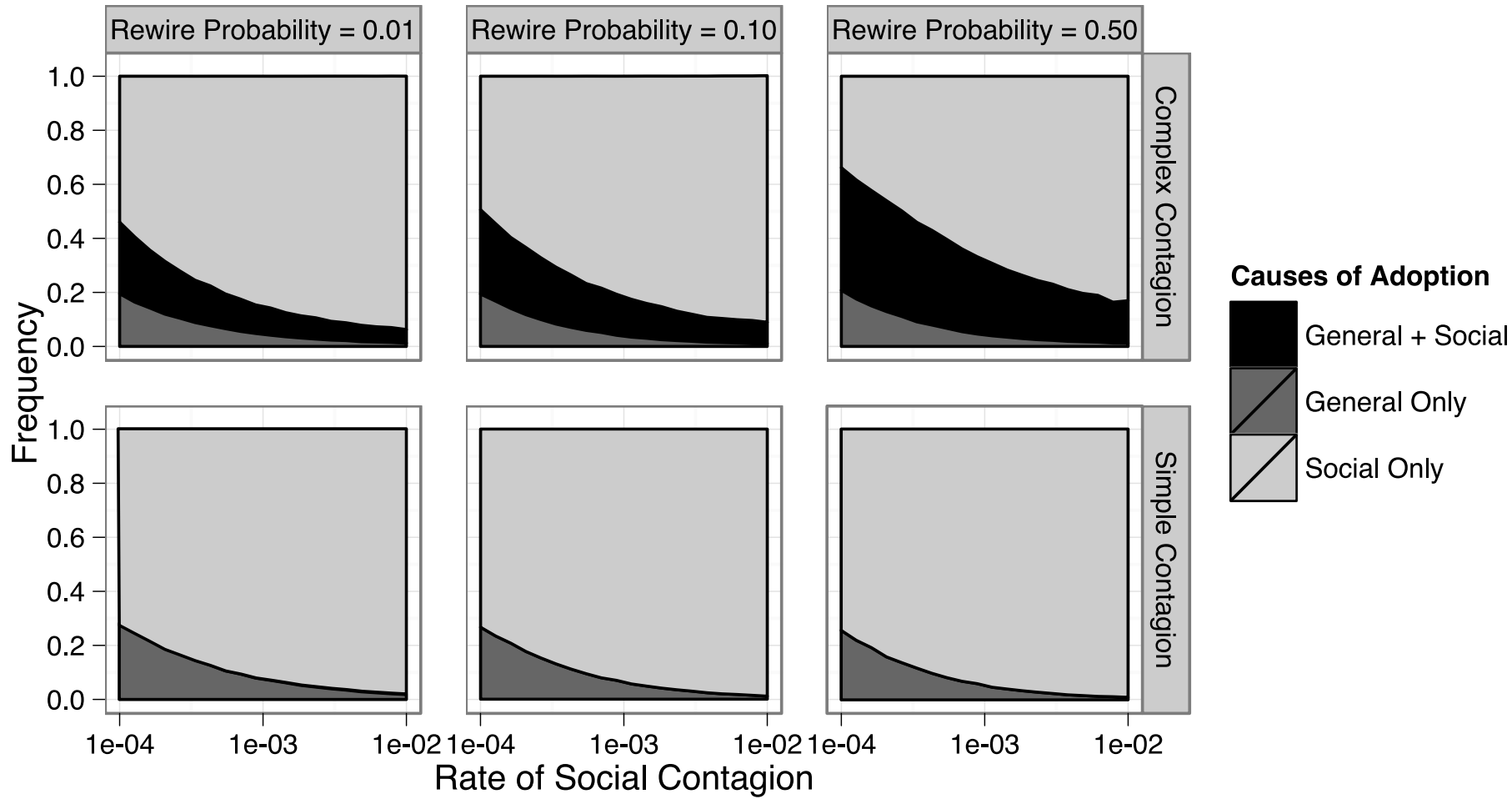

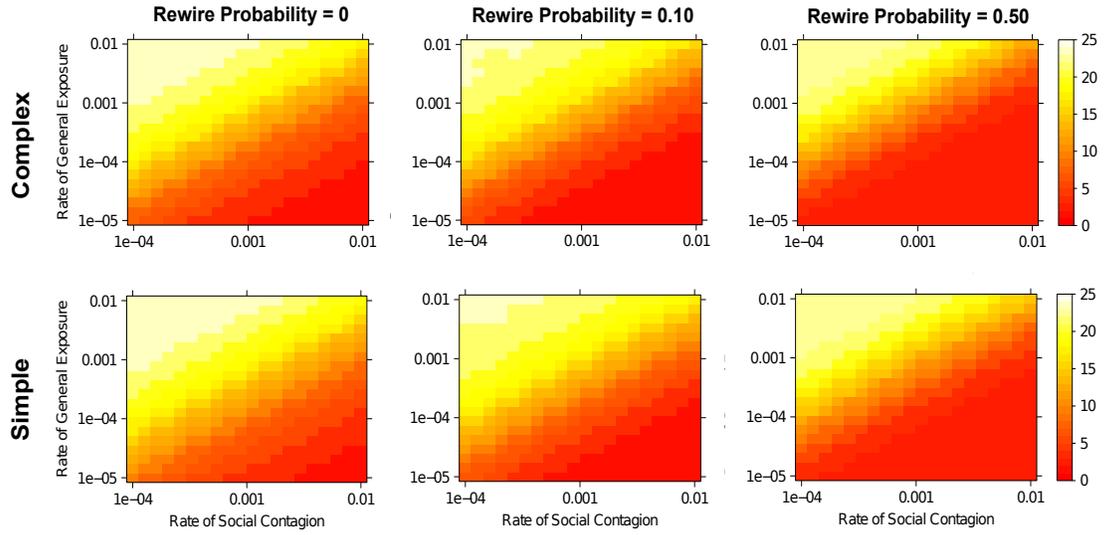